# Receding Horizon Optimization for Disturbance-aware Predictive Control of Power Electronic Inverters

Zhengxi Chen, Xun Shen, *Member, IEEE*

*Abstract*—A disturbance-aware predictive control policy is proposed for DC-AC power inverters with the receding horizon optimization approach. First a discrete event-driven hybrid automaton model has been constructed for the nonlinear inverter system dynamics. A control problem of infinite discrete state-space transition sequence optimization is formulated. A receding horizon optimization approach is applied to solve the discrete optimization problem piece-wisely on-line. Accordingly, disturbance-aware adaptive control is proposed, the external disturbance is sampled and estimated by an on-line Recursive Least Square (RLS) algorithm. Then it is elaborated that the conventional PWM control solution is a subset of solutions of the proposed control strategy and the code-transition between them is provided. By adding extra PWM constraints to the proposed control strategy, an Optimal PWM Control Mode (OPCM) is introduced as example. The proposed controller can freely operate under the original Optimal Discrete Control Mode (ODCM) and the OPCM. The numerical simulation results have verified that the proposed discrete control strategy has realized disturbance-aware adaptive control of DC-AC inversion against load-shift, and ODCM has better control performance than OPCM. In addition, the proposed modeling and control frame has potential to support other forms of control modes.

## I. INTRODUCTION

THE equivalent impedance model and state space model are widely applied for power-electronic-based inverters and converters nowadays [1]. The equivalent impedance method in the frequency domain treats the non-linear system of converters and inverters based as black boxes. Since they are expected to behave as power sources [2], linearized equivalent voltage or current source model based on Thevenin or Norton equivalence are implemented to model the electronic systems where the classic linear Nyquist stability criterion can be conveniently applied for model design and stability analysis [3]. [4] has provided a review of topologies of various impedance-source-networks-based power converters for modeling. However, the equivalent impedance method linearizes and neglects transient dynamics of the discrete switching process of controllable power electronic switches [1] [5]. On the other hand, the state-space model in the time domain, has been widely used for power system stability analysis [6]. The averaged state-space model is commonly used [7]-[9], where the transient discrete switching operation is linearized and averaged by modulation duty ratios when modeling. However, linearization is an approximation in the neighborhood of operating points and cannot predict nonlinear dynamic details [10].

Linear strategies based on PWM techniques and PID control design have been widely applied for power converters and inverters. [11] has discussed different implementation structures such as the *dq* frame and PI control structures for grid-connected power-electronic-based distributed power generations. [5] has provided a review of various control and modulation strategies for converters based on impedance model. However, the conventional PID control does not always have a clear method to define control design constants and usually requires a trial-and-error procedure [12], while the PWM techniques limits the full utilization of the switching frequency of power electronic devices. The linear models and control methods will compromise the high-level flexibility provided by the high-frequency nonlinear switching operation. Thus, many studies also have attempted to implement nonlinear control tools. [13] has implemented sliding mode control for the impedance method Z-source inverter based on PI control. [14]-[18] have applied various predictive control methods

Hybrid dynamic system is a combination of discrete events and continuous dynamic sub-systems, and hybrid automaton is a very successful method to model and analyze hybrid system [19]. The hybrid automaton modeling directly describes discrete features and provides a flexible framework for application of various linear or nonlinear control strategies and mathematic tools. For power-electronic-based converters and inverters, it can conveniently and directly describe the discrete events of "switching" and the corresponding continuous system evolution between each switching. In [20], with novelty, hybrid automaton modeling has been implemented to dual-active-bridge converters with the randomized optimization approach to solve its control problem. But, it has not 1) further explored the relations between the proposed non-linear control strategy and conventional PWM techniques based on hybrid automaton modeling; 2) proposed corresponding regulation strategies against disturbances.

This paper, based on the hybrid automaton modeling, has elaborated that the essence of the control problem for power inverters is to solve an infinite optimal discrete state-space-transition sequence. Then, a multiple-step model-predictive-control (MPC)-based discrete control strategy is proposed to solve a finite receding horizon optimization on-line, achieving piece-wise control optimality to approximate the infinite global optimal solution. Different from conventional fixed-frequency PWM methods, the proposed control strategy is frequency-varying, fully freeing the flexibility of power electronic switches. It is obvious that conventional PWM control solutions are sub-sets of solutions of the proposed control strategy. As example, an Optimal PWM Control Mode based on the proposed non-linear optimal control method is introduced by adding extra PWM constraints to the controller. The proposed controller can freely operate under the original Optimal Discrete Control Mode (ODCM) and the OPCM, or other designated modes if required.

In addition, besides methods such as following physical characteristics of synchronous machines to yield virtual inertia against disturbances [21], this paper has introduced a self-adaptive disturbance estimation method in accordance with the MPC method based on the hybrid automaton modelling. The external disturbance is treated as an unknown parameter in the system model and is estimated through data sampling, processing and an on-line estimation algorithm embedded in the controller. It is interesting to note that, the process for the controller to estimate external disturbances and offset their impacts through optimization is the self-adaptive control here. The adaptiveness comes from the "learning" performances of the estimation algorithm, i.e., the damping speed of estimation errors may resemble the control response speed.

The rest of the paper is organized as follows. Section II discusses the circuit configuration and the switching process of inverters before formulating the control problem. Section III constructs the hybrid automaton modelling of the inverter. Then the receding horizon optimization policy of control and the disturbance learning method are proposed. Later in this section, the Optimal PWM Control Mode based on the proposed control strategy is introduced. The simulation validation for the control method including Optimal Discrete Control Mode and Optimal PWM Control Mode is provided in Section IV. Finally, Section V concludes the whole study and points out the direction of future research.

## II. PROBLEM DESCRIPTION

### A. Circuit Configuration of Inverters

This paper has chosen the three-phase full bridge voltage-type inverter with an LCL filter under symmetric conditions for the control study. A resistant load is connected to the inverter through a transmission line. Regarding the given topology, shown in Fig. 1, it is feasible to control the output of each phase independently. For simplicity, the six-switch three-phase system is decoupled into three two-switch independent symmetric sub-systems of phase A, B and C with a phase angle difference of 120°.

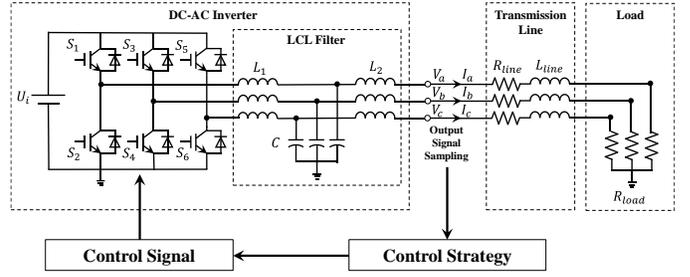

Fig. 1. The three-phase full bridge voltage inverter with an LCL filter connects the load through a transmission line.

The independent two-switch circuit structure of phase $P \in \{A, B, C\}$ is shown in Fig. 2, that

$$\begin{cases} (S_{up}, S_{down})_A = (S_1, S_2) \\ (S_{up}, S_{down})_B = (S_3, S_4) \\ (S_{up}, S_{down})_C = (S_5, S_6) \end{cases} \quad (1)$$

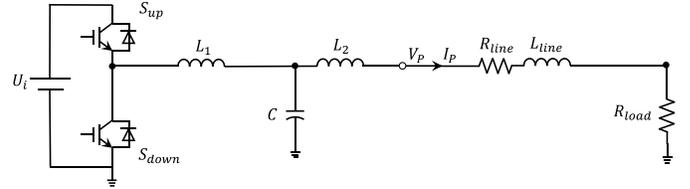

Fig. 2. The circuit structure of Phase $P$.

For each independent phase, there are totally $2^2 = 4$ operating modes based on the states (on or off) of the switches which cannot both switch to "on" to avoid short-circuit at the DC side. Hence, the switching process is described by the remaining 3 modes.

Let $M$ denote all discrete physical operating modes of the inverter and $m$ denote a certain physical state of switches, $\forall \alpha_i \in \{on, off\}, i \in [1,6]$.

$$\begin{pmatrix} M_A \\ M_B \\ M_C \end{pmatrix} = \begin{pmatrix} \cup \{(S_1(\alpha_1), S_2(\alpha_2))\} \\ \cup \{(S_3(\alpha_3), S_4(\alpha_4))\} \\ \cup \{(S_5(\alpha_5), S_6(\alpha_6))\} \end{pmatrix} = \begin{pmatrix} \cup \{m_A\} \\ \cup \{m_B\} \\ \cup \{m_C\} \end{pmatrix} \quad (2)$$

$$M = \{M_A, M_B, M_C\} = \cup \{m\} = \cup \{m_A, m_B, m_C\} \quad (3)$$

For phase $P \in \{A, B, C\}$, the combination of operating modes based on switch states can be described as following.

$$M_P = \{m_{P,1}, m_{P,2}, m_{P,3}\}$$
$$= \left\{ \begin{array}{c} \left(S_{up}(on), S_{down}(off)\right), \left(S_{up}(off), S_{down}(on)\right), \\ \left(S_{up}(off), S_{down}(off)\right) \end{array} \right\} \quad (4)$$

Under the sub-circuit structure of each discrete mode $m_{p,i}$, there exists a continuously evolving dynamic system described in form of state space equations. Under phase $P$, let $x_{P,1}, x_{P,2}, x_{P,3}$ respectively denote physical parameters $i_{P,1}, i_{P,2}, v_{P,1}$, that

$$X_P = (x_{P,1}, x_{P,2}, x_{P,3})^T. \quad (5)$$

The equivalent sub-circuit structure and corresponding state space equations $\dot{X}_P = f_P(X_P)$ of $m_{P,1}$, $m_{P,2}$ and $m_{P,3}$ are as following.

1) Mode $m_{P,1} = (S_{up}(on), S_{down}(off))$

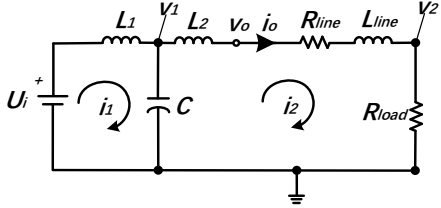

Fig. 3. Equivalent sub-circuit structure of phase $P$ under mode $m_{P,1}$.

$$f_{P,1} = \begin{pmatrix} 0 & 0 & -\frac{1}{L_1} \\ 0 & -\frac{R_{line} + R_{load}}{L_2 + L_{line}} & \frac{1}{L_2 + L_{line}} \\ \frac{1}{C} & -\frac{1}{C} & 0 \end{pmatrix} \cdot X_P + \begin{pmatrix} \frac{U_i}{L_1} \\ 0 \\ 0 \end{pmatrix} \quad (6)$$

2) Mode $m_{P,2} = (S_{up}(off), S_{down}(on))$

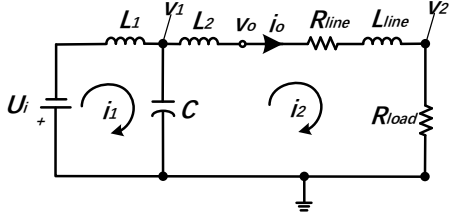

Fig. 4. Equivalent sub-circuit structure of phase $P$ under mode $m_{P,2}$.

$$f_{P,2} = \begin{pmatrix} 0 & 0 & -\frac{1}{L_1} \\ 0 & -\frac{R_{line} + R_{load}}{L_2 + L_{line}} & \frac{1}{L_2 + L_{line}} \\ \frac{1}{C} & -\frac{1}{C} & 0 \end{pmatrix} \cdot X_P + \begin{pmatrix} -\frac{U_i}{L_1} \\ 0 \\ 0 \end{pmatrix} \quad (7)$$

3) Mode $m_{P,3} = (S_{up}(off), S_{down}(off))$

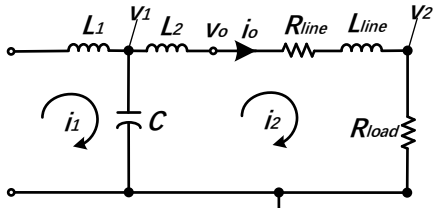

Fig. 5. Equivalent sub-circuit structure of phase P under mode $m_{P,3}$.

$$f_{P,3} = \begin{pmatrix} 0 & 0 & 0 \\ 0 & -\frac{R_{line} + R_{load}}{L_2 + L_{line}} & \frac{1}{L_2 + L_{line}} \\ \frac{1}{C} & -\frac{1}{C} & 0 \end{pmatrix} \cdot X_P$$

and $x_{P,1} = 0$ \quad (8)

In addition, the output voltage $v_{P,o}$ and current $i_{P,o}$ under phase $P$ can be obtained according to the following equations.

$$\begin{cases} v_{P,o} = v_{P,1} - L_2 \cdot \frac{di_{P,2}}{dt} = x_{P,3} - L_2 \cdot \dot{x}_{P,2} \\ v_{P,load} = v_{P,2} = x_{P,3} - (L_2 + L_{line}) \cdot \dot{x}_{P,2} - R_{line} \cdot x_{P,2} \\ i_{P,o} = i_{P,load} = x_{P,2} \end{cases} \quad (9)$$

### B. Switching Process of The Operating Inverter

Based on the circuit configuration discussed, the inverter alters the dynamic trajectories of output voltage of each phase by changing the switch-state combination. Before we designate some specific control strategy, the inverter is able to enter any one of the three modes. The switching process of output voltage of phase $P \in \{A, B, C\}$ is demonstrated in Fig. 6. The system state is switched from $m_{P,1}$ to $m_{P,2}$ at time $t_1$, and switched to $m_{P,3}$ at time $t_2$. The output trajectories evolve along $f_{P,1}$ before $t_1$, along $f_{P,2}$ from $t_1$ to $t_2$, and along $f_{P,3}$ after $t_2$.

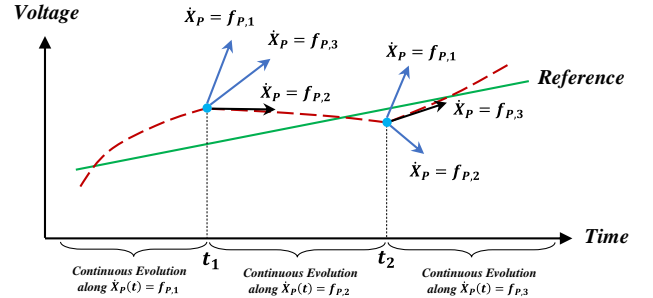

Fig. 6. The switching process of output voltage of phase $P \in \{A, B, C\}$.

The control of the inverter is the control of the switching process, including the switching order, timing and frequency. If applying the PWM control method, a possible example is to designate a control order as $m_{P,1} \to m_{P,2} \to m_{P,3}$ during every control beat, and the control frequency lower than the maximum switching frequency. Under this strategy, the switching process will follow the fixed order and the controller is expected to decide the duty ratio $d_{P,1}$, $d_{P,2}$ and $d_{P,3}$ of each mode, for phase $P \in \{A, B, C\}$. An illustration of the PWM control waveforms is shown in Fig. 7.

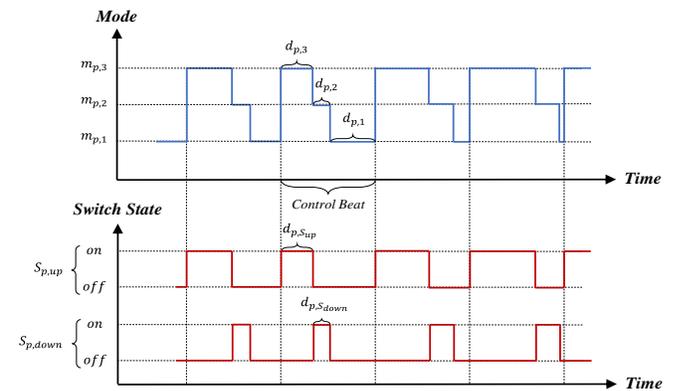

Fig. 7. Example of PWM control waveforms on phase P.

To achieve smooth adjustment of the duty ratio, the control frequency shall be much higher than the maximum switch frequency. For example, given a rated switch frequency for some power switch devices such as 100 kHz, a 50-kHz control beat frequency indicates that the duty ratio regulation can merely be 0%, 50% or 100%. In this sense, a PWM controller has to compromise its control frequency for duty ratio adjustment.

On the other hand, considering the physical switching process of an operating inverter, a controller is however at least expected to

1) have a fixed or time-varying control frequency not higher than the maximum frequency allowed;

2) decide which mode to enter or stay at every control timing.

Thus, a control strategy more flexible than the PWM method could be proposed, which does not necessarily compromise the control frequency and simply provides optimal solutions for the electronic switches to execute whenever in need. It is not hard to notice that, the PWM method solution is a sub-set of all control solutions under 1) and 2) since it has extra constraints from PWM techniques. Thus, if necessary, the potential controller would be able to imitate or realize behaviors of a PWM-method controller by adding extra constraints in its control algorithm.

### C. Control Problem Formulation

A basic task of voltage-type inverters is to invert DC voltage input into stable and reliable three-phase AC voltage output, which can be described as an optimization problem

$$\Sigma = \arg\min J(v_o, v_{ref}), \quad (10)$$

given the sinusoidal voltage reference $v_{ref}$, to obtain the optimal control $\Sigma$ minimizing the tracking error for some cost function $J(\cdot)$.

The switching process determines that system parameters will evolve along dynamic equations $f_{P,i}$ for $P \in \{A, B, C\}$ and $i \in \{1,2,3\}$. The core of control is to decide, at each control timing, into which mode the system is to be switched. Let $\sigma_P$ denote the discrete control signal and $\sigma_P = m_{P,i}$ mean that the inverter be switched into mode $m_{P,i}$. The system dynamics can be described as

$$\dot{X}_P(t) = f_P(X_P(t), \sigma_P) = f_{P,\sigma_P}(X_P(t)). \quad (11)$$

Based on any real-time system states $X_P(t)$, it is feasible to predictively calculate future system trajectories, as shown in Fig. 8. For each discrete control timing $k$, there is a discrete control $\sigma_P(k)$ which will bend the system trajectory to evolve continuously along $f_{P,\sigma_P}(X_P(t))$ from the corresponding time $t = t_k$ to the next control timing $k + 1$ with a duration of $\Delta t$. The optimization problem is hence to find an optimal discrete control solution of $\Sigma = \{\dots, \sigma_P(k), \dots\}$ for all discrete control timing that minimizes the error. Hence, *Problem* 1. becomes a following discrete predictive optimal control problem

*Problem* 1. Given the sinusoidal voltage reference $v_{ref}$, to approximate the optimal discrete control sequence $\Sigma = \{\sigma_P(k), \dots \sigma_P(k+n)\}$, $\forall n \in \mathbb{N}$, which minimize the accumulative tracking error

$$\Sigma = \bigcup_k^{k+n} \sigma_P(k) = \arg\min \int_{t_k}^{t_{k+n}+\Delta t} J\left(v_o(t), v_{ref}(t)\right) dt, \quad (12)$$

such that between each neighboring two discrete control points, the system dynamics evolve continuously along

$$\dot{X}_P(t) = f_{P,\sigma_P(k)}\left(X_P(t), \widehat{W}(t)\right),$$
$$\forall t \in [t_k, t_k + \Delta t], P \in \{A, B, C\}. \quad (13)$$

In addition, self-adaptively, the controller is expected to observe and estimate external disturbances. Based on system trajectory prediction, external disturbances will bring extra error between the prediction and real-time data samples. This error could be calculated with real-time data and used to estimate the actual disturbance $w$. The disturbance estimation $\widehat{w}$ will participate in the system trajectory prediction calculation to offset errors caused by external disturbances.

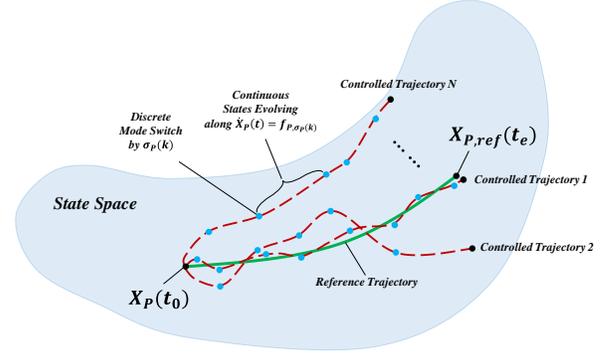

Fig. 8. The control problem is to obtain an optimal discrete control solution of $\Sigma$ which drives the system states trajectory converge to the reference.

### III. PROPOSED METHODS

### A. Hybrid Automaton Modelling

The switching behavior of a power inverter has presented a discrete event driven dynamic system. The combination of switch-states is the set of feasible discrete events, which alters the physical circuit structure and its corresponding continuous sub-system described in form of state space equations. In this sense, the switching inverter system can be well modelled by hybrid automaton from hybrid dynamic system theory. The hybrid automaton modelling method provides a framework combining both the discrete events, continuous sub-systems and their control interface. Typically, the independent hybrid automaton model for each phase of the inverter is an eight-tuple, that $\mathcal{H}_P = \{M_P, X_P, f_P, Init_P, Inv_P, E_P, G_P, R_P\}$ for $P \in \{A, B, C\}$ as shown in Fig. 9, and $\mathcal{H}_{Inverter} = \cup \{\mathcal{H}_P\}$, where

- $M_P = \{m_{P,1}, m_{P,2}, m_{P,3}\}$ is the finite set of discrete modes;

- $X_P = \{x_{P,1}, x_{P,2}, x_{P,3}\} \in \mathbb{R}^3$ represents the state space where continuous state variables evolve;

- $f_P: M_P \times X_P = \{f_{P,1}, f_{P,2}, f_{P,3}\} \to \mathbb{R}^3$, $f_{P,i}(m_{Pi}, X_P) = \dot{X}_P$, for integer $i \in \{1,2,3\}$, assigns to each discrete mode an analytic and continuous vector field $f_{P,i}$;

- $Init_P = (m_P(0), X_P(0)) \subseteq M_P \times X_P = \{m_{P,1}, m_{P,2}, m_{P,3}\} \times \{X_P \in \mathbb{R}^3 | x_{P,1} = 0, x_{P,2} = 0, x_{P,3} = 0\}$ is the initial condition;

- $Inv_P(m_{Pi}) = \{X_P \in \mathbb{R}^3 | X_P \notin G_{P,ij}\}$, $i \neq j \in \{1,2,3\}$ is the invariant set of discrete mode $m_{p,i}$ where $X_P$ keeps evolving along $f_{P,i}(m_{P,i}, X_P)$ until it reaches a guard $G_{P,ij}$, $i \neq j \in \{1,2,3\}$;

- $E_P = \cup\{(m_{P,i}, m_{P,j})\}$, $\forall i, j \in \{1,2,3\}$ is the set of all possible non-blocking executions;

- $G_P = \cup\{(m_{P,i}, m_{P,j})\}$, $\forall i, j \in \{1,2,3\}$ is the set of transition guards of all discrete modes, noting that $i = j$ means a self-transition of a mode $m_{P,i}$;

- $R_P((m_{P,i}, m_{P,3}), x_{P,1}) = 0$, $\forall i \in \{1,2,3\}$, is the forced reset mapping of the specific state $x_{P,1}$ to zero in mode $m_{P,3}$, otherwise, the value of state $X_P$ has no reset operation or is reset to itself.

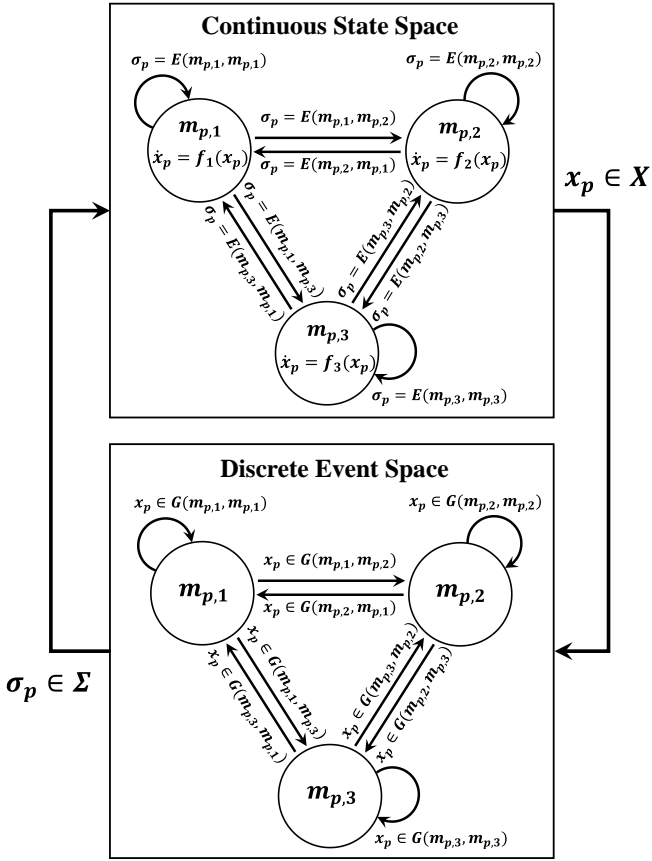

Fig. 9. The hybrid automaton model for phase $P \in \{A, B, C\}$.

TABLE I
SWITCH MODES AND SYSTEM STATE-SPACE MODELS

| Mode | Configuration of switches $(S_{up}, S_{down})$ | State-space model |
|---|---|---|
| $m_{P,1}$ | $(on, off)$ | $\dot{X}_P = f_{P,1}(X_P) = Q \cdot X_P + U$ |
| $m_{P,2}$ | $(off, on)$ | $\dot{X}_P = f_{P,2}(X_P) = Q \cdot X_P - U$ |
| $m_{P,3}$ | $(off, off)$ | $\dot{X}_P = f_{P,3}(X_P) = Q \cdot X_P$ <br> $x_{P,1} = 0$ |

where $Q = \begin{pmatrix} 0 & 0 & 0 \\ 0 & -\frac{R_{line}+R_{load}}{L_2+L_{line}} & \frac{1}{L_2+L_{line}} \\ \frac{1}{C} & -\frac{1}{C} & 0 \end{pmatrix}$, $U = \begin{pmatrix} \frac{U_i}{L_1} \\ 0 \\ 0 \end{pmatrix}$, $P \in \{A, B, C\}$

### B. Receding Horizon Optimization Policy

The operation of power inverters does not guarantee pre-set end-time. It is difficult to find the overall optimal solution for all $t \in [t_0, \infty)$. Thus, to solve *Problem* 1., this paper has adopted the receding horizon optimization policy. The controller will repeatedly estimate a piece-wise optimal solution on-line. At each control timing $k$, an optimal discrete control sequence $\{\sigma_P(k)\}$ for the future *N*-step $k \in [k, ..., k + N - 1]$ will be selected, which minimizes the *N*-step accumulative error of the output voltage. And, only $\sigma_P(k)$ for the present control timing $k$ will be executed. The optimization problem $\forall P \in \{A, B, C\}$ and $N \in \mathbb{N}^+$ can be written as

$$\sigma_P(k) = \arg \min_{\sigma_P \in \Delta_{P,M}} \int_{t_k}^{t_k + N \cdot \Delta t} (v_{P,o}(t) - v_{P,ref}(t))^2 dt, \quad (14)$$

such that system dynamics evolve continuously along

$$\dot{X}_P(t) = f_{P,\sigma_P(k+n)}\left(X_P(t), \hat{w}_{P,R_{load}}(t)\right),$$
$$\forall t \in [t_{k+n}, t_{k+n} + \Delta t], n \in [0, N-1] \subset \mathbb{N}, \quad (15)$$

within the feasible control execution set

$$\Delta_{P,M} = \cup \{E(m_{P,i}, m_{P,j})\}, \forall i, j \in \{1,2,3\}. \quad (16)$$

To solve the above optimization problem numerically, we denote the controller's frequency and period as $f_c$ and $\Delta t_c = 1/f_c$, set the numerical solver's frequency and period as $f_{sol} \geq f_c$ (integral multiple) and $\Delta t_{sol} = 1/f_{sol}$. The optimization problem for numerical solution $\forall P \in \{A, B, C\}, N \in \mathbb{N}^+$ is as

$$\sigma_P(k) = \arg \min_{\sigma_P \in \Delta_{P,M}} \left\| \hat{V}_{P,o} - V_{P,ref} \right\|^2, \quad (17)$$

where the reference value set is

$$V_{P,ref} = \left(v_{P,ref}(t_k + \Delta t_{sol}), ..., v_{P,ref}\left(t_k + N \cdot \frac{f_{sol}}{f_c} \cdot \Delta t_{sol}\right)\right), \quad (18)$$

the estimated output voltage value set is

$$\hat{V}_{P,o} = \left(\hat{v}_{P,o}(t_k + \Delta t_{sol}), ..., \hat{v}_{P,o}\left(t_k + N \cdot \frac{f_{sol}}{f_c} \cdot \Delta t_{sol}\right)\right). \quad (19)$$

For simplicity, $\forall P \in \{A, B, C\}, N \in \mathbb{N}^+$, denote the discrete control execution as

$$\sigma_P(k) = E_k(m_{P,i}, m_{P,j}) = j, \quad (20)$$

and the $N$-step discrete control sequence is

$$\Sigma_N = \{\sigma_P(k), \ldots, \sigma_P(k+N-1)\} = \{j_1, \ldots, j_N\},$$
$$\forall j_1, \ldots, j_N \in \{1, 2, 3\}. \quad (21)$$

Based on the present system values $X_P(t_k)$ at control timing $k$, $\|\hat{V}_{P,o} - V_{P,ref}\|^2$ of all $\Sigma_N$ (including $3^N$ combinations) will be calculated and compared. The sequence with the minimum value will be chosen to generate $\sigma_P(k)$. The receding horizon optimization algorithm is summarized in Table II.

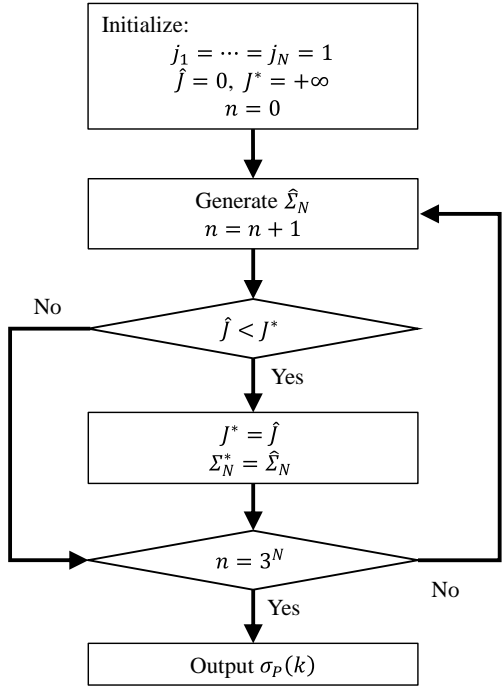

Fig. 10. The flowchart of the proposed algorithm.

TABLE II
RECEDING HORIZON OPTIMIZATION ALGORITHM

| | |
|---|---|
| At every control timing $k$, **do** | |
| Step 1: | Initialization: $$\hat{J} = 0$$ $$J^* = +\infty$$ |
| Step 2: | $N$-step discrete control sequence generation: <br>• Through traversal algorithm, generate $$\hat{\Sigma}_N = \{j_1, \ldots, j_N\}, \forall j_1, \ldots, j_N \in \{1, 2, 3\}$$ <br>• Set the counting index $$n = n + 1$$ |
| Step 3: | Based on $\hat{\Sigma}_N$, <br>• Calculate $\hat{V}_{P,o}$ <br>• Calculate $\hat{J} = \|\hat{V}_{P,o} - V_{P,ref}\|^2$ |
| Step 4: | Compare $\hat{J}$ and $J^*$, <br>• If $\hat{J} < J^*$, set $$J^* = \hat{J}$$ $$\Sigma_N^* = \hat{\Sigma}_N$$ <br>• Otherwise go to Step 5 |
| Step 5: | Output: <br>• If $n = 3^N$, output $$\sigma_P(k) = \Sigma_N^*(1)$$ <br>• Otherwise go back to Step 2 |

## C. Disturbance Learning

Many control strategies for electronic inverters and converters aim at following external characteristics of synchronous machines to realize virtual inertia against disturbances.

In this paper, self-adaptively, the controller is designed to observe and estimate external disturbances through system data samples, and offsets impacts of disturbances by feeding estimation results in the optimization computation. In this sense, the adaptiveness does not have specific physical meaning, but relies on the "learning" capability of the estimator.

The Recursive Least Square (RLS) algorithm is applied to estimate the disturbance in a time series, which provides recursive disturbance estimation for the RHO control strategy. Let the estimation result be an $N$-element vector $\hat{W} = (\hat{w}(t+1), \ldots, \hat{w}(t+N))$, the learning sample be an $m$-element vector ($m \geq N$), that means to estimate $N$ values based on $m$ history values in a time series.

TABLE III
RLS ALGORITHM FOR DISTURBANCE LEARNING

| | |
|---|---|
| Step 1: | Initialization of parameters: $$\lambda \in (0,1]$$ $$r = \vec{0}_{N \times m}$$ $$P = \delta \times \tilde{I}_{m \times m}$$ |
| Step 2: | Loading data: $$u(t) = (w(t-m-N+1), \ldots, w(t-N))^T$$ $$d(t) = (w(t-N+1), \ldots, w(t))^T$$ $$u_e(t) = (w(t-m+1), \ldots, w(t))^T$$ $$w(t) = 0, \forall t < 0$$ |
| Step 3: | Recursive calculation: <br>• The recursive calculation starts when $t = m + N$; <br>• Calculation of parameters $$\pi = u^T(t) \cdot P(t-1)$$ $$\gamma = \lambda + \pi \cdot u(t)$$ $$k(t) = \pi^T/\gamma$$ $$\alpha(t) = d(t) - r(t-1) \cdot u(t)$$ <br>• Update of parameters $$r(t) = r(t-1) + \alpha(t) \cdot k^T(t)$$ $$P(t) = [P(t-1) - k(t) \cdot \pi]/\lambda$$ <br>• Estimation output $$\hat{w}(t) = r(t) \cdot u_e(t)$$ |

In this paper, load-shifting is the chosen external disturbance for study. The disturbance sample is calculated based on the real-time samples of the system states which is

$$X(t) = \{x_1, x_2, x_3\}. \quad (22)$$

At timing $k$, the controller will estimate an optimal control signal $\sigma_{t_{k+1}}$ for the next timing $k+1$ with the estimated disturbance $\hat{W}(t_{k+1})$ added in $\mathcal{H}_P$. As the recursive on-line estimation proceeds, the actual $W(t_{k+1})$ is needed for revising estimation error and updating the profile of $\hat{W}$ sample set. $W(t_{k+1})$ will be sampled as following two steps.

1) Given $\sigma_{t_{k+1}}$, the disturbance estimator firstly estimates and stores system states of $t_{k+1}$ along $\mathcal{H}_P$, assuming no external disturbance, that

$$\hat{X}_{\sigma_{t_{k+1}}}(t_{k+1}) = \{\hat{x}_{1,\sigma_{t_{k+1}}}, \hat{x}_{2,\sigma_{t_{k+1}}}, \hat{x}_{3,\sigma_{t_{k+1}}}\}; \quad (23)$$

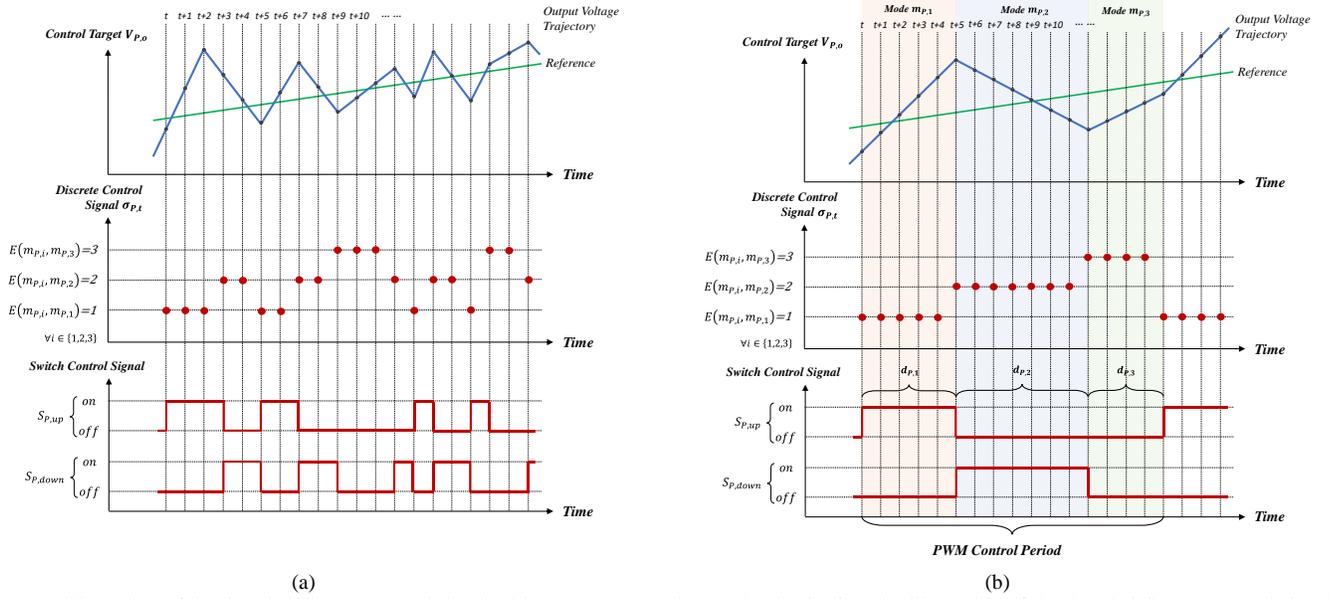

(a) (b)
Fig. 11. (a) Illustration of the decoded discrete control signal without extra constraints under the ODCM; (b) illustration of the decoded discrete control signal under the designated OPCM.

2) Then, $\sigma_{t_{k+1}}$ is executed and yields the real-time $X(t_{k+1})$. The error between $X(t_{k+1})$ and $\hat{X}_{\sigma_{t_{k+1}}}(t_{k+1})$ reflects external disturbance, and the designated load-shift disturbance could be sampled as

$$w_{R_{load}}(t_{k+1}) = R_{load}(t_{k+1}) - \hat{R}_{load}(t_{k+1})$$
$$= \frac{v_{load}(t_{k+1})}{i_{load}(t_{k+1})} - \frac{\hat{v}_{load,\sigma_{t_{k+1}}}(t_{k+1})}{\hat{i}_{load,\sigma_{t_{k+1}}}(t_{k+1})}, \quad (24)$$

where $v_{load}$ and $i_{load}$ are calculated according to Formula (7) in Section II.

It is worth mentioning that $i_{load}$ is sinusoidal. The calculation of disturbance sample will fail when $i_{load}$ crosses zero. However, the zero-crossing sampling issue is another topic not to be discussed here. In this paper, in a neighborhood of zero that $i_{load} \in [-\varepsilon, \varepsilon]$ (A) with a small $\varepsilon \in \mathbb{R}^+$, $w_{R_{load}}$ samples are set equal to the last effective sample before the current exits the zero-crossing range.

D. Decoding of The Discrete Control Signal and Optimal PWM Control Imitation

Execution $E(m_{P,i}, m_{P,j})$ means that the system is switched to the latter mode $m_{P,j}$ from $m_{P,i}, \forall i,j \in \{1,2,3\}$. If $i = j$, the system remains in the same mode. Thus, the discrete control signal can be coded as following

$$\begin{cases} E(m_{P,i}, m_{P,1}) = 1 \\ E(m_{P,i}, m_{P,2}) = 2, \\ E(m_{P,i}, m_{P,3}) = 3 \end{cases} \quad (25)$$

$$\sigma_{P,t} \in \Delta_{P,M} = \cup \{E(m_{P,i}, m_{P,j})\} = \{1,2,3\},$$
$$\forall i,j \in \{1,2,3\}, P \in \{A,B,C\}. \quad (26)$$

Without extra constraints, the optimization solution will only focus on minimizing $\|\hat{V}_{P,o} - V_{P,ref}\|^2$. As shown in Fig. 11 (a), the discrete control signal will not necessarily follow other regulations while minimizing the error. If we convert the discrete control signal to switch control signal, irregularity can be observed in the waveforms. Generally, the controller enables higher and not-fixed operation frequency that every single control timing can be fully used to activate all modes. For convenience, we denote the above-mentioned control mode as Optimal Discrete Control Mode (ODCM).

In addition, if necessary, by adding extra constraints, the controller is able to imitate other feasible control methods. In this paper, we let the controller subject to extra constraints from one possible PWM control method as example:

1) a fixed PWM control frequency $f_{PWM}$ is set;
2) the system will enter each mode in a fixed sequence of $m_{P,1} \to m_{P,2} \to m_{P,3}$ within every PWM control period;
3) the operation of switches is guided by duty ratios of three modes in each period.

For convenience, we denote the designated PWM control method as Optimal PWM Control Mode (OPCM). In Fig. 11 (b), the illustration of the controller working under the OPCM is shown, where the receding horizon optimization policy is still applied for solution. Set a PWM control frequency $f_{PWM} < f_c$ that $f_c$ is multiple times of $f_{PWM}$, and $T_{PWM} = 1/f_{PWM}$, the total number of discrete control timings in each PWM period will be $f_c/f_{PWM}$ with a duration of $\Delta t_c$. Each control beat will account for a minimum unit duty ratio of $\Delta t_c/T_{PWM}$ (%).

Denote $d_{P,i}(k)$ as the duty ratio of mode $i$ and phase $P, \forall i \in \{1,2,3\}$ and $P \in \{A,B,C\}$. Given $f_c$ and $f_{PWM}$, $\forall n_{d_1} \in [0, f_c/f_{PWM}] \subset \mathbb{N}$, $n_{d_2} \in [0, f_c/f_{PWM} - n_{d_1}] \subset \mathbb{N}$, the duty ratios of three modes subject to

$$\begin{cases} d_{P,1}(k) = n_{d_1} \cdot \frac{\Delta t_c}{T_{PWM}} \\ d_{P,2}(k) = n_{d_2} \cdot \frac{\Delta t_c}{T_{PWM}} \\ d_{P,3}(k) = 1 - d_{P,1}(k) - d_{P,1}(k) \end{cases} \quad (27)$$

Hence, the optimization problem under the OPCM for numerical solution can be rewritten as formulas (28). The solution is a set of duty ratios for future $N$- PWM periods, or $\forall n \in [0, N-1] \subset \mathbb{N}$

$$D_P = \bigcup_{k}^{k+n} D_P(k) = \bigcup_{k}^{k+n}\{(d_{P,1}(k), d_{P,2}(k), d_{P,3}(k))\}. \quad (28)$$

Similarly, only $D_P(k)$ for discrete control timing $k$ will be executed. Hence,

$$D_P(k) = \arg\min \left\| \hat{V}_{P,o} - V_{P,ref} \right\|^2. \quad (29)$$

$\forall n \in [0, N-1] \subset \mathbb{N}$, $i \in \{1,2,3\}$, $P \in \{A, B, C\}$, and duty ratios subject to

$$\begin{cases} d_{P,1}(k+n) = n_{d_1}(k+n) \cdot \dfrac{\Delta t_c}{T_{PWM}} \\ d_{P,2}(k+n) = n_{d_2}(k+n) \cdot \dfrac{\Delta t_c}{T_{PWM}} \\ d_{P,3}(k+n) = 1 - d_{P,1}(k+n) - d_{P,1}(k+n) \end{cases}, \quad (30)$$

$\forall n_{d_1} \in \left[0, \dfrac{f_c}{f_{PWM}}\right] \subset \mathbb{N}, n_{d_2} \in \left[0, \dfrac{f_c}{f_{PWM}} - n_{d_1}\right] \subset \mathbb{N},$

The system dynamics evolve continuously along

$$\dot{X}_P(t) = \begin{cases} f_{P,m_{P,1}}, t \in [t_{k+n}, t_{k+n} + d_{P,1} \cdot T_{PWM}) \\ f_{P,m_{P,2}}, t \in [t_{k+n} + d_{P,1} \cdot T_{PWM}, t_{k+n} + (d_{P,1}+d_{P,2}) \cdot T_{PWM}) \\ f_{P,m_{P,3}}, t \in [t_{k+n} + (d_{P,1}+d_{P,2}) \cdot T_{PWM}, t_{k+n} + T_{PWM}) \end{cases}$$

(31)

in the pre-designated order of $m_{P,1} \to m_{P,2} \to m_{P,3}$.

## IV. SIMULATION VALIDATION

Simulation validation in the SIMULINK environment of MATLAB is provided to verify the proposed method. The simulation starts at $t = 0.0\ sec$ and ends at $t = 1.0\ sec$. Main simulation parameters are shown in Table III.

### A. Control Performance

The comparison of the three-phase output voltage profiles of both the Optimal Discrete Control Mode (ODCM) and the Optimal PWM Control Mode (OPCM) are provided in Fig. 12, under
- Case 1: no load-shift disturbance;
- Case 2: load-shift disturbance but no RLS estimation;
- Case 3: both load-shift disturbance and RLS estimation.

TABLE III
SIMULATION PARAMETERS

| | | |
|---|---|---|
| System | (1) Line Parameters | $R_{line} = 0.5\ \Omega$ <br> $L_{line} = 0.0015\ H$ |
| | (2) LCL Filter Parameters | $L_1 = 0.009\ H$ <br> $L_2 = 0.003\ H$ <br> $C = 60 \times 10^{-6}\ F$ |
| | (3) Load | $R_{load} = 100\ \Omega$ |
| | (4) DC Source | $U_i = 800\ V$ |
| Controller | (5) Control Targets | Frequency $f_{ref} = 50\ Hz$ <br> Voltage Magnitude $V_{ref} = 220\ V$ <br> Initial Phase Angle of Phase A $\theta_{A-ref} = 30°$ <br> (Phase A, B and C has a phase angle difference of 120°.) |
| | (6) Predictive Controller | Rated Switch Frequency $f_S = 20\ kHz$ <br> ODCM Frequency $f_{ODCM} = f_S = 20\ kHz$ <br> Prediction Steps $N_{ODCM} = 5$ <br> OPCM Frequency $f_{PWM} = 4\ kHz$ <br> Prediction PWM-periods $N_{PWM} = 1$ |
| Disturbance | (7) Load-shift | $R_w = \begin{cases} +50\% \times R_{load}, & t \in [0.02\ sec, 0.05\ sec) \\ -50\% \times R_{load}, & t \in [0.05\ sec, 0.08\ sec) \\ 0, & others \end{cases}$ ($\Omega$) |
| RLS Estimation | (8) Estimation Output Size | ODCM: $N_e = N_{ODCM} = 5$ <br> OPCM: $N_e = N_{PWM} \cdot f_{ODCM}/f_{PWM} = 5$ |
| | (9) Sampling | Input Sample Set Size $m = 10$ <br> Sampling Frequency $f_e = f_{ODCM} = 20\ kHz$ |
| | (10) Initialization of Parameters | $\lambda = 1, r = \vec{0}_{N_e \times m}, P = 10^{-3} \times \vec{I}_{N_e \times m}\ (\delta = 10^{-3})$ |

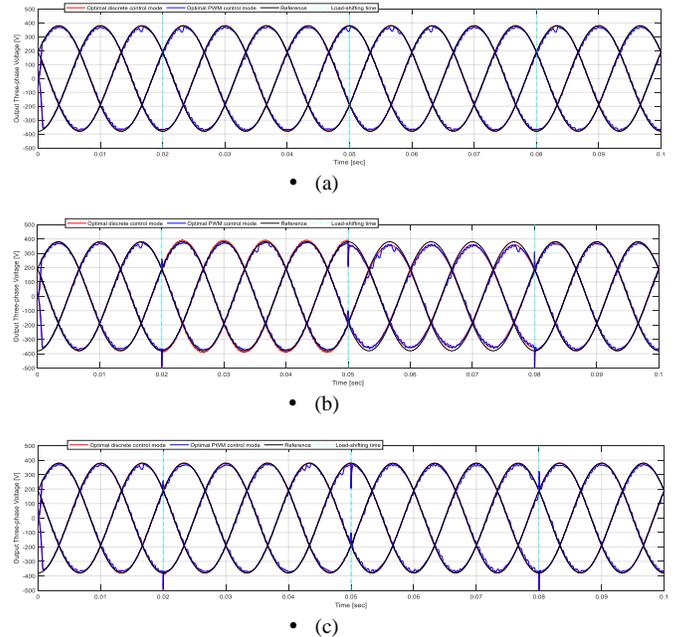

(a)

(b)

(c)

Fig.12. The output three-phase voltage of (a) Case 1; (b) Case 2; (c) Case 3.

TABLE IV
SIMULATION RESULTS OF TRACKING ERRORS

| Items | Case | Optimal discrete control mode (ODCM) | | | Optimal PWM control mode (OPCM) | | |
|---|---|---|---|---|---|---|---|
| | | Phase A | Phase B | Phase C | Phase A | Phase B | Phase C |
| Initial Value (V) | 1 | 190.0000 | 190.0000 | 380.0000 | 190.0000 | 190.0000 | 380.0000 |
| | 2 | | | | | | |
| | 3 | | | | | | |
| Settling Time (ms) | 1 | 0.62 | 0.45 | 0.77 | 0.62 | 0.46 | 0.77 |
| | 2 | | | | | | |
| | 3 | | | | | | |
| Mean Value (V) | 1 | 1.5084 | 1.5563 | 1.5923 | 8.0874 | 8.0167 | 7.9830 |
| | 2 | 7.1239 | 7.1770 | 7.1171 | 10.2093 | 11.4887 | 10.4961 |
| | 3 | 1.7990 | 1.8054 | 1.8965 | 8.5321 | 8.6145 | 8.7163 |
| Max or Overshoot Value (V) | 1 | 9.6501 | 8.6815 | 6.3912 | 46.3791 | 44.1780 | 42.6482 |
| | 2 Overshoot at $t=0.02\ sec$ | 58.0696 | 63.4053 | 127.3557 | 47.7512 | 69.6863 | 110.4293 |
| | 2 Overshoot at $t=0.05\ sec$ | 81.5986 | 81.3131 | 162.7742 | 86.9995 | 88.2332 | 172.4407 |
| | 2 Overshoot at $t=0.08\ sec$ | 99.2069 | 110.7178 | 216.5100 | 90.6754 | 119.4196 | 213.3630 |
| | 3 Overshoot at $t=0.02\ sec$ | 58.0696 | 63.4053 | 127.3557 | 47.7512 | 69.6863 | 110.4293 |
| | 3 Overshoot at $t=0.05\ sec$ | 83.6690 | 85.4501 | 167.8435 | 88.1280 | 85.8727 | 174.0024 |
| | 3 Overshoot at $t=0.08\ sec$ | 121.3637 | 125.3862 | 245.4091 | 115.1177 | 132.1363 | 229.1404 |
| Standard Deviation (V) | 1 | 2.0194 | 1.9065 | 1.8623 | 51.7897 | 48.3409 | 40.6309 |
| | 2 | 53.2242 | 54.9614 | 70.7988 | 79.8160 | 122.4241 | 97.8614 |
| | 3 | 9.5934 | 9.3057 | 31.3292 | 57.4092 | 55.0445 | 70.4814 |

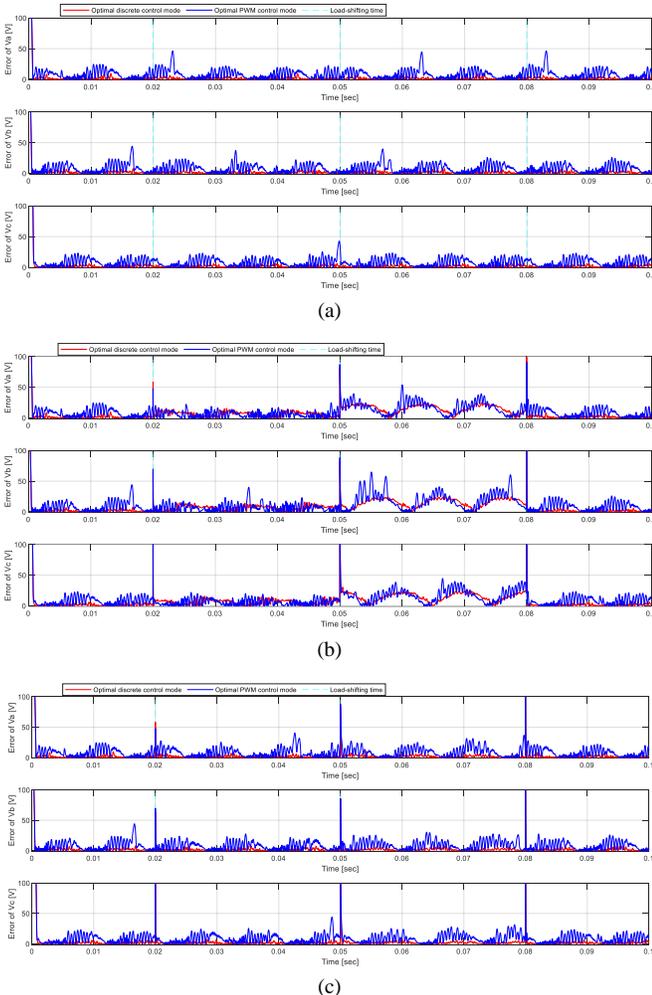

Fig.13. The tracking error of phase A, B and C under (a) Case 1; (b) Case 2; (c) Case 3.

The tracking errors are plotted in Fig. 13 and shown in Table IV. At the beginning, ODCM and OPCM have close response behavior that the initial error is damped during nearly the same settling time. Both control strategies have computed the same optimal control results for phase A and B to stay in $m_1$, phase C in $m_2$ during the initial error damping for output voltage to catch up with the reference at the fastest speed.

During the dynamic steady state, ODCM has smaller and stabler tracking errors than OPCM for all Case 1,2 and 3. When load-shift disturbance is triggered in Case 2 and Case 3, the tracking accuracy is compromised than that of Case 1 under both ODCM and OPCM.

Under Case 2, compared with Case 1, the mean errors of all phases shoot up to about 7.1-7.2 $V$ from 1.5-1.6 $V$ for ODCM, and up to 10.2-11.5 $V$ from 7.9-8.1 $V$ for OPCM. The standard deviations increase from 1.8-2 $V$ to 53-71 $V$ for ODCM, and from 40-52 $V$ to 79-122 $V$ for OPCM, that the errors of OPCM are oscillating at a larger range.

Under Case 3, the mean errors are close to Case 1 which are slightly increased to 1.7-1.9 $V$ for ODCM, and to 8.5-8.7 $V$ for OPCM. Its standard deviations are also smaller than that of Case 2, within 9-32 $V$ for ODCM and 41-68 $V$ for OPCM. It is because the RLS estimation has provided the controller with adaptiveness against the load-shift.

Simulation results have verified that ODCM has better tracking performance than OPCM due to higher computation and control frequency. However, OPCM is quite acceptable under situations not in demand of higher accuracy or in lack of capability of higher-frequency computation, that it exchanges some control accuracy with much less computation tasks and lower switching frequency. In addition, the control method proposed fully supports implementation of both or even more control modes at the same time in its frame if necessary.

## B. Input Control Signal Waveforms

### 1) Optimal Discrete Control Mode (ODCM)

The discrete control signal waveforms under ODCM of Case 1 are shown as example in Fig. 14.

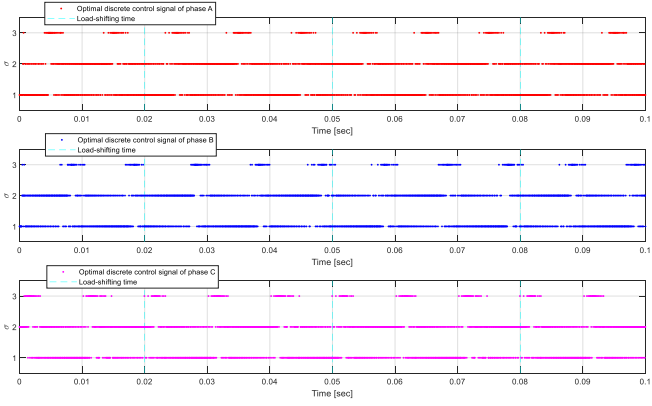

Fig.14. Waveforms of the discrete control signal $\sigma$ of phase A, B, C.

As discussed in Section III, ODCM is designed to focus on solutions of the optimization problem depicted in Formula (17) at the designated frequency of 20 $Hz$ which is 5 times that of OPCM. Thus, it might not be possible and necessary to convert the discrete signals into waveforms with particular regularities such as PWM waveforms. Here, to demonstrate the relation between discrete signals and the switching process in a direct manner, the discrete signals are decoded into corresponding switch-states $\{(S_1, S_2), (S_3, S_4), (S_5, S_6)\}$ according to

$$(S_1, S_2) = \begin{cases} (1,0), \text{if } \sigma_A = 1 \\ (0,1), \text{if } \sigma_A = 2, \\ (0,0), \text{if } \sigma_A = 3 \end{cases} \quad (32)$$

$$(S_3, S_4) = \begin{cases} (1,0), \text{if } \sigma_B = 1 \\ (0,1), \text{if } \sigma_B = 2, \\ (0,0), \text{if } \sigma_B = 3 \end{cases} \quad (33)$$

$$(S_5, S_6) = \begin{cases} (1,0), \text{if } \sigma_C = 1 \\ (0,1), \text{if } \sigma_C = 2. \\ (0,0), \text{if } \sigma_C = 3 \end{cases} \quad (34)$$

In Fig. 15, details of the discrete control signal $\{\sigma_A, \sigma_B, \sigma_C\}$ and corresponding switch-states $\{(S_1, S_2), (S_3, S_4), (S_5, S_6)\}$ for $t \in [0.043, 0.044]$ (sec) are shown.

### 2) Optimal PWM Control Mode (OPCM)

The duty ratio waveforms under OPCM through simulation of Case 1 are shown as example in Fig. 16. Under OPCM, the discrete optimal predictive controller follows the pre-designated PWM control rules and solve the corresponding optimization problem depicted from Formula (28) to (31), so that the controller imitates the PWM control mechanism. However, it is worth mentioning that the core of computation is still based on solving a discrete receding horizon optimization problem, but with extra constraints. Thus, the duty ratios for PWM control shown in Fig. 16 are yielded through decoding original solutions in the form of discrete control signals.

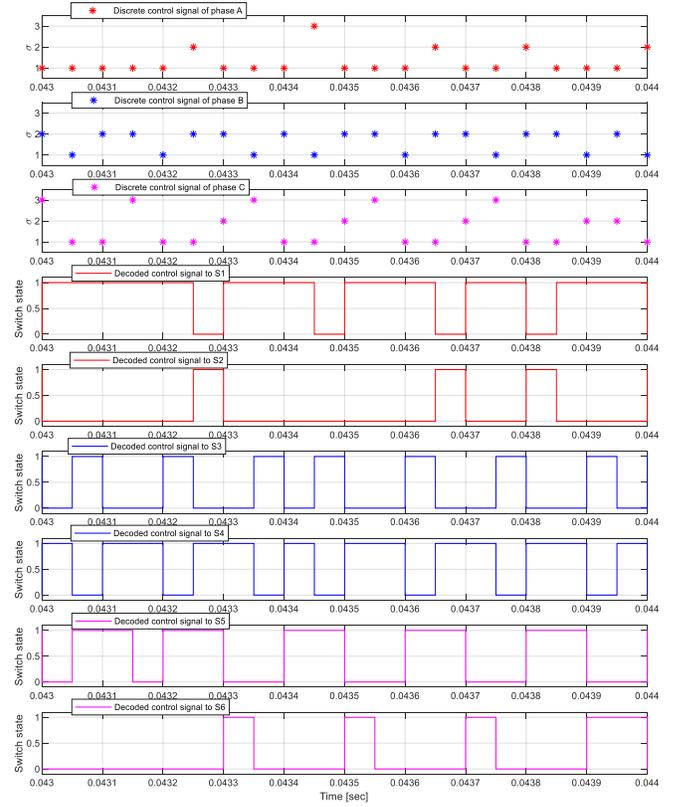

Fig. 15. Details of discrete control signals and corresponding switch-states for $t \in [0.043, 0.044]$ (sec).

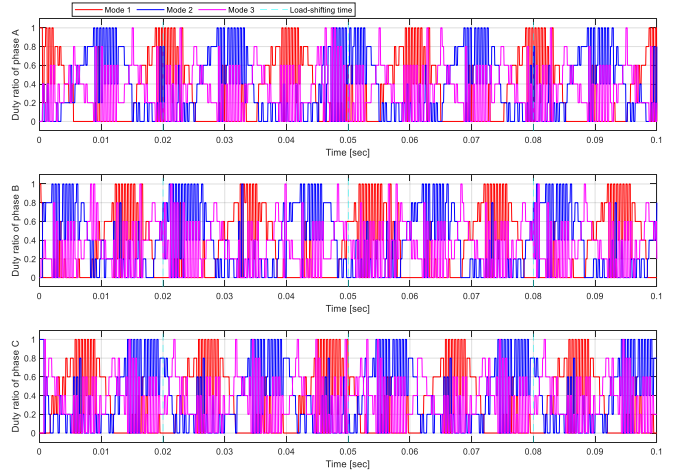

Fig.16. Waveforms of the duty ratios for phase A, B, C.

To demonstrate the decoding process, in Fig. 17, details of original discrete control signals $\{\sigma_A, \sigma_B, \sigma_C\}$ and corresponding PWM duty ratios $\{d_{P,1}, d_{P,2}, d_{P,3}\}_{P=A,B,C}$ after decoding for $t \in [0.043, 0.044]$ (sec), are plotted. It is shown that, the discrete control signals are no longer irregular, but following the PWM control constraints under the pre-designated mode transition sequence of $m_1 \to m_2 \to m_3$ by regulating respective duty ratios at the PWM frequency of 4 $Hz$.

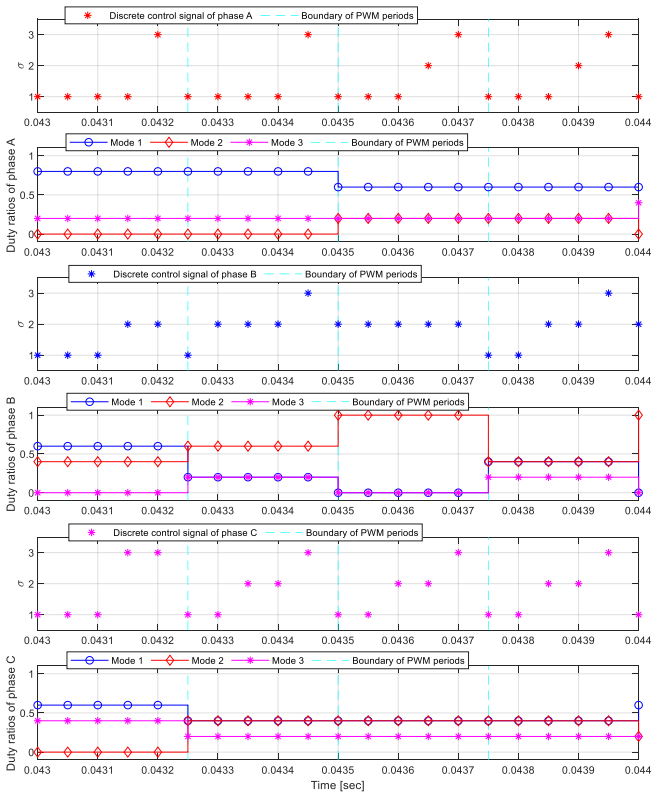

Fig.17. Details of discrete control signals and corresponding PWM duty ratios after decoding for $t \in [0.043, 0.044]$ (sec).

## C. RLS Estimation

The RLS estimation simulation result is shown in Fig. 18. At $t = 0.02\ sec$, $0.05\ sec$ and $0.08\ sec$, when the step-function load-shift occurs, a respective estimation error overshoot appears at the magnitude of 50 Ω, -23.44 Ω and 3.41 Ω before being damped.

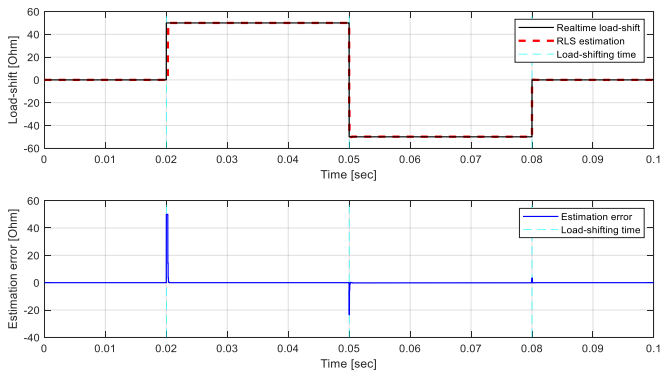

Fig.18. The RLS estimation result and estimation errors and details of error overshoots at $t \in \{0.02, 0.05, 0.08\}$ (sec).

## V. CONCLUSION AND FUTURE WORK

This paper has proposed an adaptive predictive control policy through receding horizon optimization for DC-AC inverters based on hybrid automaton modelling. The adaptiveness is obtained through application of RLS algorithm which provides the controller with estimation of external disturbances based on real-time samples of system parameters. The proposed control framework also supports imitation of PWM control solutions by adding extra control constraints. Numerical simulations have been implemented for validation. The main contribution of the proposed method has the following three aspects.

1) The proposed control method fully utilizes the rated physical switching frequency of power electronics and has smaller tracking errors. From simulation, the average tracking errors of Optimal Discrete Control Mode (ODCM) are reduced by 78% to 81% compared to the Optimal PWM Control Mode (OPCM) which follows extra PWM control constraints. Even when external load-shift disturbance occurs without adaptive mechanisms, ODCM's tracking errors are still 30% to 38% smaller.
2) The nonlinear and discrete modelling framework allows further application of machine learning theories, that for example, the reinforcement learning method can be implemented to map quick solutions to on-line controller based on recorded off-line data under general operation conditions, which will reduce computation burdens.
3) The PWM control solutions are subsets of all solutions of the proposed method. Thus, by adding extra control constraints, the proposed control method can switch to other operation modes which imitate or approximate PWM control strategies if necessary.
4) The adaptiveness in this paper does not necessarily obtain specific physical meaning, but is in accordance with the "learning" performances of the disturbance data sampling and estimation algorithm.

Since the proposed method has only been numerically simulated, the future work is to conduct experimental validation.

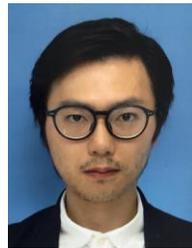
**XUN SHEN** (M'19) was born in Yueyang, China. He received the B.S. degree in electrical engineering from Wuhan University, Wuhan, China, in 2012, and the Ph.D. degree in mechanical engineering from Sophia University, Tokyo, Japan, in 2018. He is currently pursuing the Ph.D. degree with the Department of Statistical Sciences, The Graduate University for Advanced Studies. He has been a Postdoctoral Research Associate with the Department of Engineering and Applied Sciences, Sophia University, since April 2018. Since June 2019, he has been an Assistant Professor with the Department of Mechanical Systems Engineering, Tokyo University of Agriculture and Technology. His current research interests include chance constrained optimization, point process, and their applications in power systems and intelligent vehicles.

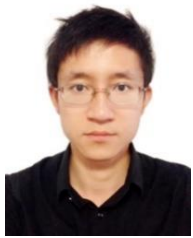
**ZHENGXI CHEN** was born in Wuhan, China. He received the B.S. degree in electrical engineering from Wuhan University, Wuhan, China, in 2012, and M.S. degree in electrical engineering from Washington University in Saint Louis, Saint Louis, United States, in 2014. From 2014 to 2017, he worked at State Grid Economic and Technological Research Institute CO., LTD. From 2017 to now, he is working at Global Energy Interconnection Group Corporation Limited of State Grid Corporation of China. His current research interests include hybrid dynamic system, optimal optimization, machine learning and their application in power system.